\documentclass[11pt,twoside]{article}

\usepackage{asp2014}

\aspSuppressVolSlug
\resetcounters

\def\schwa{\rotatebox[origin=c]{180}{e}}

\begin{document}

\title{A data model to connect the ESO Data Processing System (EDPS) to ELT data archives}

\author{Hugo Buddelmeijer$^{1,2}$, and Gijs Verdoes Kleijn$^1$}
\affil{$^1$Kapteyn Astronomical Institute, University of Groningen, 9747 AD, Groningen, The Netherlands; \email{buddel@astro.rug.nl}}
\affil{$^2$Institute for Astrophysics, University of Vienna, 1180 Wien, Austria}

\paperauthor{Buddelmeijer~H.}{buddel@astro.rug.nl}{0000-0001-8001-0089}{University of Groningen}{Kapteyn Astronomical Institute}{Groningen}{}{}{Netherlands}
\paperauthor{Verdoes~Kleijn~G.~A.}{g.a.verdoes.kleijn@rug.nl}{0000-0001-5803-2580}{University of Groningen}{Kapteyn Astronomical Institute}{Groningen}{}{}{Netherlands}

\begin{abstract}
We extend the data-driven approach for astronomical pipelines to the METIS instrument and integrate the ESO Data Processing System (EDPS) by adapting it to support backward chaining.
\end{abstract}

\section{Ambition}
To fully harness the scientific potential of our advanced telescopes, we advocate for a holistic approach to astronomical pipelines.
This involves seamlessly integrating data, the code that processes it, the workflow that generates it, and its visualization, ensuring each component enhances the others.

A pivotal aspect of our approach is the adoption of a backward-chaining data model \citep{2013ExA....35..131M}, where workflows are constructed starting from the desired end product.
This allows the workflow engine to optimize the process by knowing the final goal from the outset.

We outline our steps toward this vision in the context of the archive for the Mid-infrared ELT Imager and Spectrograph (METIS) \citep{2021Msngr.182...22B} that will be used during the instrument integration, and test phase.

\section{METIS}
METIS is a first-generation instrument for the Extremely Large Telescope (ELT), Europe's upcoming 39-meter optical and infrared telescope.
It offers diffraction-limited imaging, low- and medium-resolution slit spectroscopy, coronagraphy for high-contrast imaging from 3 to 13 microns, and high-resolution integral field spectroscopy from 3 to 5 microns.
METIS's main scientific goals include the detection and characterization of exoplanets, the study of protoplanetary disks and planet formation, stellar formation and evolution, the galactic center, active galactic nuclei, and our solar system.

Following the Final Design Review (FDR) of METIS in the fall of 2022, the project entered the Manufacturing, Assembly, Integration, and Testing (MAIT) phase\footnote{See \url{https://elt.eso.org/instrument/METIS/} for the latest status of METIS.}.
During this phase, raw dataproducts coming from the Instrument Control System and reduced science dataproducts coming from the datareduction/pipeline computer are stored in a dedicated METIS AIT Archive.

\section{METIS Data Model}
We build on the approach developed for MICADO \citep{2020ASPC..527..127B}, combining WISE-technology with a declarative data model.
WISE-technology, initially created for the Kilo Degree Survey, features an Object Relational Mapper with a LINQ-like query syntax for Python, while remaining language-agnostic for database interactions.
This approach has been adapted for various languages, including XML variants for Euclid, MUSE, and MICADO.

For METIS, we introduced a simpler method: directly parsing the data model from the pipeline design document.
The METIS Data Reduction Library Design document, written in LaTeX, follows the updated structure required by ESO for the ELT.
A parser ensures the document's internal consistency and generates a machine-readable data model for the archive\footnote{Software for this conversion are available on demand}.

The LaTeX content is parsed into Python dataclass instances, which are then converted into Python files to create database tables and columns.
Each data item type, such as a Raw Dark or a Master Flat, corresponds to a Python class and a database view, with FITS file headers mapped to database columns.

This comprehensive data model allows for client development in various languages, although Python is used for convenience.
Astropy \citep{2018AJ....156..123A} facilitates interaction with the file based representation of the data, supporting formats beyond FITS files.

\section{Implicit Workflows}
A data reduction library can be divided into:
\begin{itemize}
\item Data Model: Description of each data item, its properties, and file representation.
\item Task Definitions: The logic and processing code for creating data items using other data items as input.
\item Workflow Rules: The grouping and association of data items.
\end{itemize}


\noindent A workflow engine performs two actions to process a specific set of data:
\begin{enumerate}
  \item Generating a set of data processing jobs from Task Definitions and Workflow Rules.
  \item Executing these jobs to process the data.
\end{enumerate}
This paper focuses on the first action, constructing a graph of interdependent jobs to achieve the desired target data product.
The execution of these tasks is beyond the scope of this paper.

Our core proposition is that there is no need for explicit task definitions and workflow rules, because they can be subsumed in the data model.
That is, we believe a complete set of jobs can be constructed straight from the data model if one ensures the data model does not only specify the data items, but also the relationships between different types of input data items.
All the required information can be encoded as local relationships; there is no need to have separate Workflow Rules \footnote{We do however, also provide a manually created workflow for the METIS pipeline.}.

Past attempts at deriving the workflow directly from the data model had difficulties expressing certain workflow structures \citep{2013A&A...559A..96F}.
However, we argue that the limitations experienced by those projects were caused by the workflow engine, and can be resolved by using what we call a backward-chaining workflow engine (Figure~\ref{chaining}).

\section{Backward Chaining}
In a forward-chaining system, the workflow engine takes the raw data, and uses the Workflow Rules and Task Definitions to determine what it can make from that.
That is, workflow engine interprets the Task Definition into jobs starting with the tasks closest to the raw data.
This is turned around in a backward-chaining system: the workflow engine starts interpreting with the task that creates a desired end product, and recursively works backwards towards the raw data.

In typical scenarios, both approaches result in exactly the same graph of tasks to create the exact same data product from the input data.
The difference becomes clear once information needs to be propagated from one task to another.

A forward chaining system starts with all tasks `disconnected' from each other.
That is, the tasks that are interpreted first are not using each other's output as input; they all have raw data as input.
The entire graph of tasks only becomes connected once the last task is interpreted into the job that will create the desired target data product.
It is necessary to have an overarching set of workflow rules to connect the disconnected graph to move information around.

For example, if interpreting a specific task leads to the conclusion that a certain calibration data product (like a specific Master Flat) should always be preferred over other similar data products, then that information needs to be propagated to other tasks that would otherwise select another data product.

The graph of tasks is always connected in a backward-chaining system.
It is therefore trivial to propagate information over the edges of the graph; either by pushing it through when created, or even better, pulling it when needed.

\articlefigure{C701_f1.eps}{chaining}{
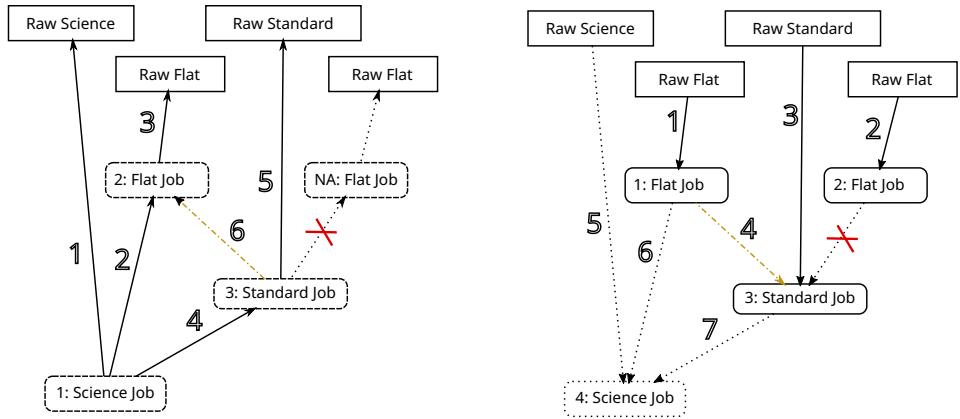
  Backward (left) and forward (right) chaining; numbers represent the order in which the associations and jobs are created.
  Usually, the raw science and the raw standard observation should be processed each with their own master flat; that is, the crossed out association is the default. 
  However, in this particular scenario, the standard star reduction should use the same flat as is used for the science data reduction.
  That is, the orange dashed association should be created instead of the crossed out one.
  Creating the correct association is easy in the backward-chaining figure, because the relevant parts of the dependency graph are already connected.
  However, it is hard in the forward-chaining figure, because the graph consists of three disconnected parts, because the dotted associations have not yet been formed.}

\section{ESO Data Processing System (EDPS)}
The EDPS \citep{2024A&A...681A..93F}, ESO's new workflow engine, replaces several older systems.
Backward compatibility with the earlier workflow engines is a priority, so it primarily uses forward chaining.

We have developed SPDE (stylized as sdp\schwa), a fork of EDPS with backward chaining support.
Contact the authors for access to the SPDE pre-release.

\section{Discussion}
The METIS AIT Archive will be created using a data model derived from the design document.
Data processing will be performed using both the standard EDPS and the backward-chaining SPDE for comparison to verify they are equivalent.
Future enhancements for a data-centric backward-chaining system include interleaving job creation and execution, dynamic dependency graph modifications, and lazy/partial job execution.
This approach promises simpler workflows and better optimization of the final data product, making such systems indispensable in the future.


\bibliography{C701}  


\end{document}